\documentstyle[aps,prd,eqsecnum,epsfig]{revtex}
\begin{document}
\draft

\thispagestyle{empty}

\title{Stability criterion for self-similar solutions 
with perfect fluids in general relativity}
\author{Tomohiro Harada \footnote{Electronic
 address:harada@gravity.phys.waseda.ac.jp}}   
\address{Department of Physics, Waseda University, Shinjuku, Tokyo
169-8555, Japan}  
%
%\date{\today}
\maketitle

\begin{abstract}                % DON'T CHANGE THIS LINE
A stability criterion is derived for self-similar solutions 
with perfect fluids which obey the equation of state $P=k\rho$ 
in general relativity.
A wide class of self-similar solutions
turn out to be unstable against the so-called kink mode. 
The criterion is directly related
to the classification of sonic points.
The criterion gives a sufficient condition for instability of the solution. 
For a transonic point in collapse, 
all primary-direction nodal-point solutions are unstable,
while all secondary-direction nodal-point 
solutions and saddle-point ones are stable against the kink mode.
The situation is reversed in expansion.
Applications are the following:
the expanding flat Friedmann solution for
$1/3 \le k < 1$ and the collapsing one for
$0< k \le 1/3$ are unstable; 
the static self-similar solution is unstable;
nonanalytic self-similar collapse solutions
are unstable;
the Larson-Penston (attractor) solution
is stable for this mode for $0<k\alt 0.036$, while
it is unstable for $0.036\alt k $;
the Evans-Coleman (critical) solution is stable 
for this mode for $0<k\alt 0.89$, while it is unstable for $0.89\alt k$.
The last application suggests that the Evans-Coleman solution
for $0.89\alt k $ is {\em not critical} because it has 
at least two unstable modes. 
\end{abstract}
\pacs{PACS numbers: 0420D, 0425D, 0440, 0440N}

\section{introduction}

The invariance under scale transformation
is one of the important features of gravitational physics
as long-range force 
in general relativity as well as in Newtonian gravity.
The scale invariance of field equations 
implies the existence
of solutions which are invariant under the scale transformation.
Such scale-free solutions are called 
{\em self-similar solutions}.
Among them, the most widely researched system in general relativity 
is a spherically symmetric
self-similar spacetime with a perfect fluid,
which is pioneered by Cahill and Taub~\cite{ct1971}.
The existence of such solutions 
requires the equation of state to be the form $P=k\rho$
if it is barotropic.
A wide class of matter fields can be modelled 
by this equation of state.
In general relativity, spherically symmetric 
self-similar solutions are classified 
with the theory of dynamical systems~\cite{gnu1998} 
and 
with their asymptotic forms~\cite{cc2000}. 
For a recent review, see~\cite{cc1999}.

Although self-similar solutions form a zero-measure set of 
solutions to the field equations,
several authors conjectured that self-similar solutions
play important roles in gravitational collapse and/or
in cosmological situations (see, e.g.,~\cite{carr1999}).
In this context, the so-called Larson-Penston self-similar solution
has been shown to describe generic gravitational collapse
of isothermal gas in Newtonian 
gravity~\cite{hn1997hm2000,mh2001,ti1999}.
This solution was discovered by Penston and independently 
by Larson~\cite{lp1969},
and after that generalized to general relativity 
by Ori and Piran~\cite{op1987,op1990}.
Actually, Newtonian self-similar solutions are 
obtained by taking the limit $k\to 0$ of those 
in general relativity~\cite{op1990}.
Recently, it has been shown
by numerical simulations and a normal mode analysis
that the general relativistic counterpart of the Larson-Penston
solution is an {\em attractor solution} 
in general relativistic gravitational collapse~\cite{hm2001}. 
This convergence phenomena, i.e., 
the generic convergence to a self-similar solution, 
can be regarded as the recovery of scale-invariance
symmetry in gravitational physics because
the scale-invariance symmetry
which the system of the field equations originally has
is regained in gravitational collapse even if the symmetry is 
lost in the initial situation. 
In the Larson-Penston solution, 
a naked singularity develops from regular initial data 
for $0<k\alt 0.0105$~\cite{op1987,op1990}.
See also~\cite{hm2001,harada1998}
for the relationship between 
the cosmic censorship conjecture and the nature of 
the Larson-Penston solution as an attractor.

In addition to the attractive nature of a self-similar solution, 
the critical behavior in the gravitational
collapse discovered by Choptuik~\cite{choptuik1993}
shed light on self-similarity (for a recent 
review, see~\cite{gundlach1998}).
He investigated the threshold between the collapse to a black hole
and the dispersion to infinity
in the self-gravitating system of a massless scalar field.
He found critical behavior,
such as the mass-scaling law for the formed black hole mass,
which is analogous to that in statistical physics.
He also found that there is a discrete self-similar solution 
that sits
at the threshold of the black hole formation, which is called 
a {\em critical solution}.
After that, Evans and Coleman~\cite{ec1994} observed similar 
phenomena in the collapse 
of a radiation fluid ($k=1/3$)
and found that the critical solution coincides with one of the 
continuous self-similar solutions with analyticity, which
we call Evans-Coleman solution~\footnote{
Although this self-similar solution itself had been already 
discovered by Ori and Piran~\cite{op1990}, 
we call this solution, together with its generalization
to more general values of $k$,
Evans-Coleman solution for convenience in this paper.}.
Recently, their work was extended to $0<k\le 1$~\cite{nc2000}.
A renormalization group approach by Koike et al.~\cite{kha1995}
gave a simple physical explanation to the critical phenomena and 
showed that the critical solution is characterized by a single
unstable mode for a radiation fluid.
After that their work was extended to 
$0<k\le 0.889$~\cite{maison1996kha1999}.
The detailed characteristics of the Evans-Coleman solution
for $0<k\le 1$ was investigated~\cite{nc2000,ccgnu2000}.
Moreover,
`Hunter (a)' self-similar solution,
which was originally discovered by Hunter~\cite{hunter1977}, 
was identified with the Newtonian counterpart of 
the Evans-Coleman solution~\cite{mh2001,hm2001}.

For Newtonian self-similar solutions,
Ori and Piran~\cite{op1988} pointed out the existence of
{\em kink instability}
in a wide class of self-similar solutions.
The kink mode results from the existence of sonic points.
The nature of sonic points was examined
in Newtonian gravity~\cite{shu1977ws1985}
and in general relativity~\cite{op1990,bh1978,cy1990,fh1993}.
The kink mode corresponds to the injection of a weak discontinuity
at sonic points at some initial moment.
The instability is characterized by the divergence
of the discontinuity in a finite time interval.
The blow-up of this mode implies shock wave formation.
Unfortunately, the analysis was restricted to Newtonian 
gravity and applies at most to 
the case $0<k\ll 1$~\cite{op1990}.
In this paper, the analysis is generalized to fully 
general relativistic case.
The stability criterion obtained here
applies even to the case for any value of $k$ ($0<k<1$).
Then, we can deal with highly relativistic situations,
such as the collapse of a radiation fluid,
the evolution of early universe, and so on.
The obtained criterion gives a sufficient condition for
instability and a necessary condition for stability,
because the analysis is specified on the kink mode. 
Of course, the present analysis reproduces
the previously obtained results in Newtonian gravity.

The organization of this paper is the following.
In Section II, basic equations are presented.
In Section III, the classification of sonic points are
reviewed, which is closely related to the 
kink instability.
In Section IV, the stability for the kink mode is analyzed
in full order, and a stability criterion for this mode
is derived.
In Section V, applications to known 
self-similar solutions, such as
the flat Friedmann solution, 
static self-similar solution,
Larson-Penston solution and Evans-Coleman solution,
are presented.
In Section VI, we comment on recent numerical works by 
Neilsen and Choptuik~\cite{nc2000} and discuss
two interesting limiting cases.
Section VII is devoted to summary. 
We adopt units such that $G=c=1$. 
%%---------------------------%
\section{Basic Equations}
%%---------------------------%
\subsection{Einstein's Equations}
In this section, we basically follow 
the formulation on a spherically 
symmetric self-similar system given by Bicknell and Henriksen~\cite{bh1978}.
It is noted that there are several different 
formulations 
(see, e.g.,~\cite{gnu1998,cc2000,op1990}).

The line element in a spherically symmetric spacetime is given by
\begin{equation}
ds^{2}=-e^{\sigma(t,r)}dt^{2}+e^{\omega(t,r)}dr^{2}+R^{2}(t,r)(d\theta^{2}
+\sin^{2}\theta d\phi^{2}).
\end{equation}
We consider a perfect fluid as a matter field
\begin{equation}
T^{\mu\nu}=(\rho+P)u^{\mu}u^{\nu}+Pg^{\mu\nu}.
\end{equation}
Here we adopt the comoving coordinates.
Then Einstein's equations and the equations of motion for 
the matter field are reduced to the following 
partially differential equations (PDE's):
\begin{eqnarray}
& &\frac{\partial \sigma}{\partial r}=-\frac{2}{\rho+P}
\frac{\partial P}{\partial r}, 
\label{eq:dsigmadr}\\
& &\frac{\partial \omega}{\partial t}=-\frac{2}{\rho+P}
\frac{\partial \rho}{\partial t}-\frac{4}{R}\frac{\partial R}
{\partial t}, 
\label{eq:domegadt}\\
& &\frac{\partial m}{\partial r}=4\pi R^{2}\rho\frac{\partial R}
{\partial r}, 
\label{eq:dmdr}\\
& &\frac{\partial m}{\partial t}=-4\pi R^{2} P\frac{\partial R}
{\partial t}, \\
& &e^{-\sigma}\left(\frac{\partial^{2}R}{\partial t^{2}}
-\frac{1}{2}\frac{\partial\sigma}{\partial t}\frac{\partial R}{\partial t}
\right)-\frac{1}{2}e^{-\omega}\frac{\partial \sigma}{\partial r}
\frac{\partial R}{\partial r}=-\frac{m}{R^{2}}-4\pi PR,
\label{eq:euler}
\end{eqnarray}
and
\begin{equation}
m=\frac{R}{2}\left[1+e^{-\sigma}\left(\frac{\partial R}
{\partial t}\right)^{2}-e^{-\omega}\left(\frac{
\partial R}{\partial r}\right)^{2}\right],
\label{eq:m}
\end{equation}
where $m(t,r)$ is called Misner-Sharp mass 
and five of the above six equations are independent.

We assume the following equation of state:
\begin{equation}
P=k\rho,
\end{equation}
where we assume $0<k<1$.
We define dimensionless functions such as
\begin{eqnarray}
\eta&\equiv &8\pi r^{2}\rho,\\
S&\equiv&\frac{R}{r}, \\
M&\equiv &\frac{2m}{r}.
\end{eqnarray}
We also define the following self-similar coordinate:
\begin{equation}
\xi\equiv  \frac{t}{r}.
\end{equation}
It is often more convenient to use the following coordinates:
\begin{eqnarray}
\tau &\equiv & -\ln |t|, \\
z &\equiv& -\ln|\xi|.
\end{eqnarray}
It is noted that
$\xi$ is a two-valued function of $z$,
such as $z=-\infty $ corresponds to $\xi=\pm\infty$ 
and $z=+\infty $ corresponds to $\xi=\pm 0$.

It is found that equations (\ref{eq:dsigmadr}) and 
(\ref{eq:domegadt}) are integrated as
\begin{eqnarray}
e^{\sigma}&=&a_{\sigma}(t)
(\eta e^{-2z}) ^{-\frac{2k}{1+k}}, \\
e^{\omega}&=&a_{\omega}(r)\eta^{-\frac{2}{1+k}}S^{-4},
\end{eqnarray}
where $a_{\sigma}(t)$ and $a_{\omega}(r)$ are arbitrary functions.
These functions correspond to
the freedom of re-scaling
the time and radial coordinates as $\tilde{t}=\tilde{t}(t)$
and $\tilde{r}=\tilde{r}(t)$.
Using this freedom, we let $a_{\sigma}$ and $a_{\omega}$ be
constant.

For later convenience, we define a quantity $y$ as
\begin{equation}
y\equiv\frac{M}{\eta S^{3}},
\end{equation}
which has a clear physical meaning that it is 
one third of the ratio of the `averaged density'
of the region interior to $(t,r)$ to the local 
density at $(t,r)$.
We also define two velocity functions $V$ and $V_{R}$. 
The $V$ is the velocity of the 
$\xi=\mbox{const}$ curve relative to the fluid element,
which is written as
\begin{equation} 
V= \mbox{sign}(\xi)e^{z+\frac{\omega-\sigma}{2}},
\end{equation}
while $V_{R}$ is the velocity of the $R=\mbox{const}$ curve 
relative to the fluid element,
which is written as
\begin{equation}
V_{R}\equiv -e^{\frac{\omega-\sigma}{2}}
\frac{\left(\frac{\partial R}{\partial t}\right)}
{\left(\frac{\partial R}{\partial r}\right)}
=V\frac{\dot{S}+S^{\prime}}{S+S^{\prime}},
\label{eq:VR}
\end{equation}
where 
\begin{eqnarray}
^{\cdot} &\equiv & \frac{\partial}{\partial\tau}, \\
' &\equiv& \frac{\partial}{\partial z}.
\end{eqnarray}
$V$ is written as
\begin{equation}
V^{2}=\frac{a_{\omega}}{a_{\sigma}}(\eta e^{-z})^{-2\frac{1-k}{1+k}}S^{-4}.
\label{eq:V2}
\end{equation}

Using these dimensionless variables, 
equations~(\ref{eq:dmdr})-(\ref{eq:m}) are reduced to
the following form:
\begin{eqnarray}
& &M+M'=\eta S^{2}(S+S'), 
\label{eq:tauz1}\\
& &\dot{M}+M'= -k\eta S^{2}(\dot{S}+S'), 
\label{eq:tauz2}\\
& &\ddot{S}+2\dot{S}'+S''+\left[\frac{k}{1+k}\left(\frac{\dot{\eta}}
{\eta}+\frac{\eta'}{\eta}-2\right)+1\right](\dot{S}+S')\nonumber \\
\label{eq:tauz3}
&=&-\frac{k}{1+k}
e^{\sigma-\omega-2z}\left(\frac{\eta'}{\eta}-2\right)
(S+S')-\frac{1}{2}e^{\sigma-2z}
\frac{M+k\eta S^{3}}{S^{2}},
\label{eq:tauz4}
\end{eqnarray}
and 
\begin{equation}
\frac{M}{S}=1+e^{-\sigma+2z}(\dot{S}+S')^{2}-e^{-\omega}(S+S')^{2}, \\
\label{eq:tauz5}
\end{equation}
where three of the above four equations are independent.

\subsection{Self-similarity}
For self-similar solutions, we assume that all dimensionless
quantities depend only
on $z$, i.e., 
$M=M(z)$, 
$S=S(z)$,
$\eta=\eta(z)$, 
$\sigma=\sigma(z)$, and 
$\omega=\omega(z)$.
Then, from equations~(\ref{eq:tauz1})-(\ref{eq:tauz5}), 
we obtain the following ordinary differential equations
(ODE's):
\begin{eqnarray}
M'&=&\frac{k}{1+k}\frac{1-y}{y}M, 
\label{eq:M'} \\
S'&=&-\frac{1-y}{1+k}S, 
\label{eq:S'}\\
\eta'&=&\left[2(1-y)-2\frac{ky-\frac{1}{4}(1+k)^{2}e^{\omega}\eta}
{V^{2}-k}\right]\eta,
\label{eq:eta'}
\end{eqnarray}
and constraint equation
\begin{equation}
V^{2}(1-y)^{2}-(k+y)^{2}+(1+k)^{2}
e^{\omega}S^{-2}
(1-y\eta S^{2})=0. 
\label{eq:const}
\end{equation}
From equations~(\ref{eq:VR}) and (\ref{eq:S'}), we find 
\begin{equation}
V_{R}=-V\frac{1-y}{k+y}.
\end{equation}

\section{Sonic points}
The classification of sonic points in general relativity 
was studied by Bicknell and Henriksen~(\cite{bh1978};
see also~\cite{op1990,cy1990,fh1993}).
Here we briefly review their work,
which turn out to be closely related to
the stability criterion of the kink mode.
\subsection{Regularity}
The sonic point is defined by
\begin{equation}
V^{2}=k,
\label{eq:sonic}
\end{equation}
that is, $V$ is equal to the sound speed.
At the sonic point, the system of the ODE's 
(\ref{eq:M'})-(\ref{eq:eta'}) is singular.
Since $V$ is the relative velocity
of the $\xi=\mbox{const}$ curve to the fluid element,
informations with the sound speed 
propagate only outward or inward
for $V^{2}>k$, while in both directions
for $V^{2}<k$, in terms of $\xi$.
In this sense,
the former is supersonic, while the latter is 
subsonic, in terms of $\xi$.
Hence, informations 
can cross the sonic point only in a single direction. 
This is a physical reason why
the system of the ODE's becomes singular at the sonic point.

First, we consider the regularity condition at the sonic point.
The regularity requires
\begin{equation}
ky-\frac{1}{4}(1+k)^{2}e^{\omega}\eta=0.
\label{eq:regular}
\end{equation}
From equations~(\ref{eq:const}), (\ref{eq:sonic}) and (\ref{eq:regular}), 
we find
the following explicit relations at the sonic point
\begin{eqnarray} 
e^{-z_{s}}&=&\left[\frac{(k+y_{s})^{2}+k(3y_{s}-1)(1+y_{s})}
{a_{\sigma}^{\frac{1+3k}{2(1-k)}}
a_{\omega}^{-\frac{1}{2}}
(1+k)^{\frac{4k}{1-k}}
k^{\frac{3}{2}}
(4y_{s})^{\frac{1-3k}{1-k}}}
\right]^{\frac{1+k}{1+3k}}, \label{eq:zs}\\
M_{s}&=&y_{s}\eta_{s}S^{3}_{s}, 
\label{eq:ms} \\
S_{s}&=& 
\left[\frac{a_{\sigma}a_{\omega}(1+k)^{4}}{k(4y_{s})^{2}}\right]^{\frac{1}{4}}
e^{-\frac{1-k}{2(1+k)}z_{s}}, 
\label{eq:ss}\\
\eta_{s}&=&\left[\frac{4y_{s}}{a_{\sigma}(1+k)^{2}}\right]
^{\frac{1+k}{1-k}}e^{2z_{s}}.
\label{eq:etas}
\end{eqnarray}
Therefore, sonic points are parametrized by 
one parameter $y_{s}$.
Equation~(\ref{eq:zs}) requires the following condition: 
\begin{equation}
y_{s}>y_{1}\equiv \frac{-2k+\sqrt{k(1+6k-3k^{2})}}{1+3k}.
\label{eq:regularity}
\end{equation}

\subsection{Classification of sonic points}
\begin{figure}[htbp]
\vspace*{1cm}
\centerline{
\epsfxsize 15cm \epsfysize 17cm
\epsfbox{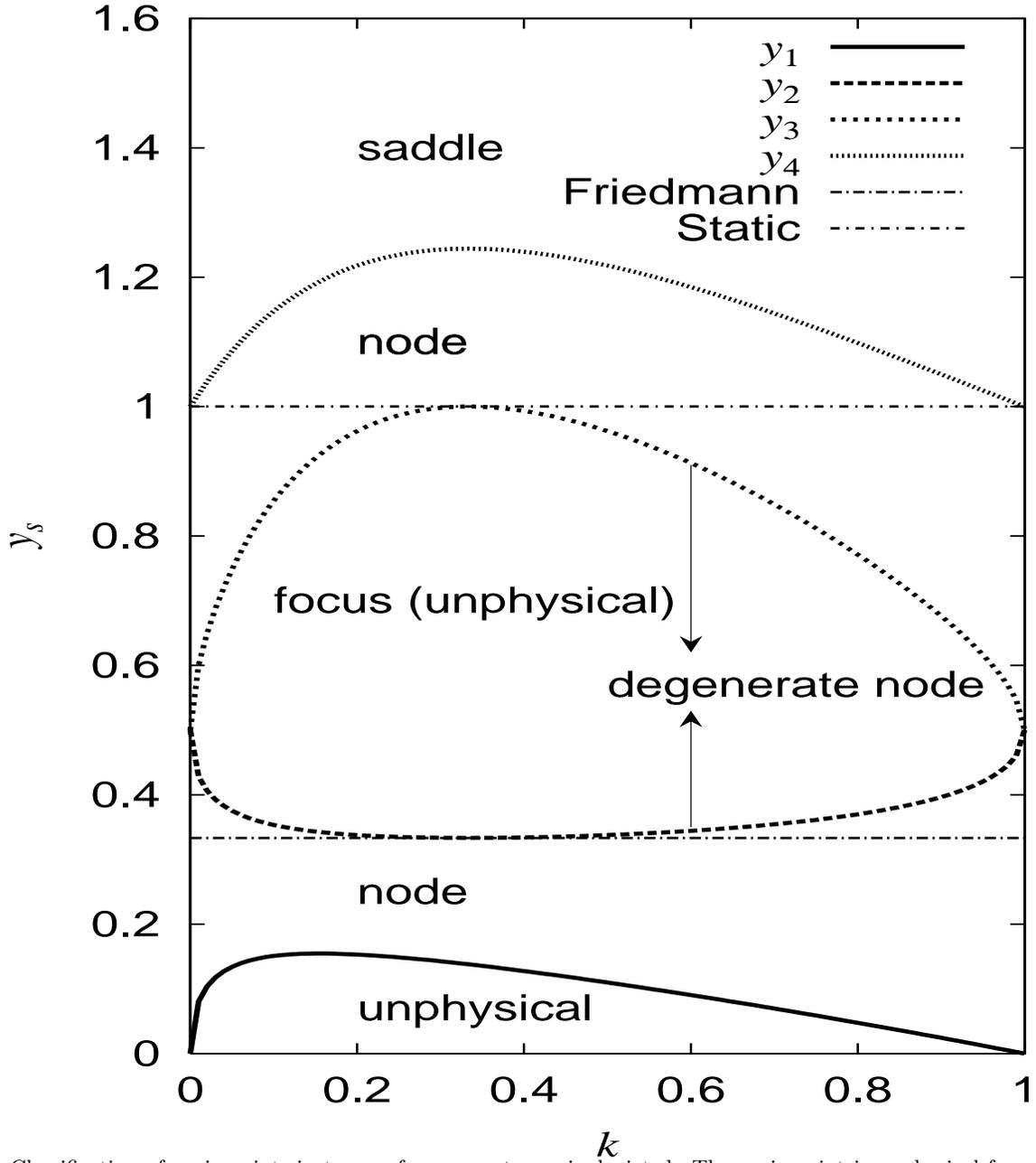}}
\caption{Classification of sonic points 
in terms of a parameter $y_{s}$ is depicted.
The sonic point is unphysical for $y_{s}<y_{1}$,
a nondegenerate node for $y_{1}<y_{s}<y_{2}$ or $y_{3}<y_{s}<y_{4}$,
a focus (unphysical) for $y_{2}<y_{s}<y_{3}$,
a degenerate node for $y_{s}=y_{2}$ or $y_{s}=y_{3}$, 
and a saddle for $y_{4} < y_{s}$. 
The flat Friedmann solution ($y_{s}=1/3$) 
and static self-similar solution ($y_{s}=1$) are also plotted.
See text for details.}
\label{fg:sonicpoint}
\end{figure}

We introduce new independent variable $u$ which is defined as
\begin{equation}
-\frac{d z}{d u}=V^{2}-k.
\end{equation} 
Using $u$ in place of $z$, 
a sonic point turns out to be a stationary point
(or a singular point) of the resultant system of the ODE's as
\begin{eqnarray}
\frac{dM}{du}&=&-\frac{k}{1+k}\frac{1-y}{y}(V^{2}-k)M, \\
\frac{dS}{du}&=&\frac{1-y}{1+k}(V^{2}-k)S, \\
\frac{d\eta}{du}&=&-\left[2(1-y)(V^{2}-k)-
2\left(ky-\frac{1}{4}(1+k)^{2}e^{\omega}\eta\right)\right]\eta,
\end{eqnarray}
A stationary point is classified by the
behavior of solutions around the point. 
We put 
\begin{eqnarray}
M&=&\eta_{s}S_{s}^{3}(y+x_{1}), \\
S&=&S_{s}(1+x_{2}), \\
\eta&=&\eta_{s}(1+x_{3}), \\
\xi&=&\xi_{s}(1+x_{4}),
\end{eqnarray}
where $x_{1}$ to $x_{4}$ are
regarded as components of the vector 
${\bf x}$.
Then, we find the following linearized 
ODE's in the matrix form:
\begin{equation}
\frac{d}{du}{\bf x}={\bf A}{\bf x},
\end{equation}
where 
the matrix ${\bf A}$ is given by
\begin{equation}
{\bf A}=\frac{2k}{1+k}\left(\begin{array}{cccc}
0 & 2k(1-y) & k(1-k)(1-y)/(1+k) & k(1-k)(1-y)/(1+k) \\
0 & -2(1-y) & -(1-k)(1-y)/(1+k) & -(1-k)(1-y)/(1+k) \\
1+k & (4-3y)(1+k) & 2(1-k-y) & 2(1-k)(1-y) \\
0 & -2(1+k) & -(1-k) & -(1-k) 
\end{array}\right).
\end{equation}
This matrix has two zero eigenvalues and two 
generically nonzero eigenvalues $\lambda_{\pm}$
($\lambda_{-}\le \lambda_{+}$) 
\begin{equation}
\lambda_{\pm}=\frac{2k}{1+k}\tilde{\lambda}_{\pm},
\end{equation}
where $\tilde{\lambda}_{\pm}$ are given by
\begin{equation}
\tilde{\lambda}_{\pm}=\frac{-(1+k)\pm\sqrt{D}}{2},
\label{eq:tildelambda}
\end{equation}
with
\begin{equation}
D=1+6k-3k^{2}-4y-24ky+12k^{2}y+4y^{2}+12ky^{2}.
\end{equation}
These two eigenvalues $\lambda_{\pm}$ 
are associated with the corresponding 
eigenvectors ${\bf e}_{\pm}$, respectively.
The sonic point is classified into a saddle
for $\lambda_{-}< 0 < \lambda_{+}$,
a nondegenerate node for $ \lambda_{-}< \lambda_{+}<0$,
a degenerate node for $\lambda_{-}=\lambda_{+}$,
and a focus for complex values of $\lambda_{\pm}$.
For a saddle, only two solutions pass through it, one along 
the $+$ive direction ${\bf e}_{+}$ and the other along the $-$ive 
direction ${\bf e}_{-}$.
For a nondegenerate node, there are one-parameter family of 
solutions which cross the sonic point along the $+$ive direction, 
while only one solution crosses along the $-$ive direction. 
The directions $+$ and $-$ for a node are called 
the primary and secondary directions, respectively. 
For a degenerate node, two directions $+$ and $-$ coincide
with each other.
The sonic point which is a focus is unphysical.
The sonic point is a nondegenerate 
node or saddle for $D>0$, a degenerate node for $D=0$,
and a focal point for $D<0$.
The condition $D\ge 0$ requires 
\begin{equation}
y_{s}\le y_{2}~\mbox{or}~y_{3} \le y_{s},
\end{equation}
where $y_{2,3}$ ($y_{2}<y_{3}$) are given by
\begin{equation}
y_{2,3}\equiv \frac{1+6k-3k^{2}\mp\sqrt{3k(1-k)(1+6k-3k^{2})}}
{2(1+3k)},
\end{equation}
and $D=0$ holds only for $y_{s}=y_{2}$ or $y_{s}=y_{3}$. 
Together with equation~(\ref{eq:regularity}), since $y_{1}<y_{2}$ can be verified,
it is found that
\begin{equation} 
y_{1}< y_{s}\le y_{2}~\mbox{or}~y_{3} \le y_{s},
\end{equation}
for the sonic point to be physical.
Whether the sonic point is a node or saddle 
is determined by the sign
of $[(1+k)^{2}-D]$.
It changes its sign at $y_{s}=y_{4}$ or $y_{s}=y_{5}$
($y_{4}>y_{5}$), where
\begin{equation}
y_{4,5}=\frac{1+6k-3k^{2}\pm\sqrt{(1+4k-k^{2})^{2}+8k^{2}(1-k)^{2}}}{2(1+3k)}.
\end{equation} 
Since $y_{5}<y_{1}$ can be verified,
it is found that the sonic point is unphysical for 
$y_{s}<y_{1}$ or $y_{2}<y_{s}<y_{3}$, 
a degenerate node for $y_{s}=y_{2}$ or $y_{s}=y_{3}$,
a nondegenerate node for $y_{1}<y_{s}<y_{2}$ or $ y_{3}<y_{s}<y_{4}$, 
and a saddle for $y_{4}< y_{s}$.
The classification of sonic points in terms of the
parameter $y_{s}$ is plotted in figure~\ref{fg:sonicpoint}.

It is seen that, along the allowed directions $\pm$,  
\begin{equation}
\frac{\eta'}{\eta}=\frac{3-k-2y+\tilde{\lambda}_{\pm}}{1-k}
\label{eq:eta's}
\end{equation}
at the sonic point.
Using this and the relation
\begin{equation}
\frac{V'}{V}=-\frac{1-k}{1+k}\frac{\eta'}{\eta}+\frac{3-k-2y}{1+k},
\label{eq:v'eta'}
\end{equation}
which follows from equations~(\ref{eq:V2}) and (\ref{eq:S'}),
it is found that
\begin{equation}
(V^{2})'=-\lambda_{\pm},
\end{equation}
or 
\begin{equation}
\frac{V'}{V}=\frac{1}{2}\mp \frac{\sqrt{D}}{2(1+k)},
\end{equation}
where and hereafter the upper and lower signs denote the
$+$ive and $-$ive directions, respectively. 
Therefore, 
if a solution crosses a saddle sonic point from subsonic interior
to supersonic exterior, then it goes along the
$-$ive direction.
A sonic point with subsonic interior and supersonic exterior
is called a `transonic' point.
If we consider a solution with regular center and sonic points, 
the solution must have at least
one transonic point.
In principle, a solution may have other kinds of sonic points,
such as the one with supersonic interior and subsonic exterior
which we call an anti-transonic point.
It is found that 
the sonic point with supersonic interior and subsonic exterior
must be a saddle and the solution passes it along the $+$ive direction.
Furthermore, it is found that $V'/V<1/2$ or $(V^{2})'<k$
for the $+$ive direction,
$V'/V>1/2$ or $(V^{2})'>k$
for the $-$ive direction, and 
$V'/V=1/2$ or $(V^{2})'=k$ for a degenerate node.

\subsection{Analyticity}
Furthermore we can require the analyticity 
condition (Taylor-series 
expandability) 
at the sonic point, i.e., 
\begin{eqnarray}
M&=&M_{s}+M_{s,1}(z-z_{s})+M_{s,2}(z-z_{s})^{2}+\cdots, \\
S&=&S_{s}+S_{s,1}(z-z_{s})+S_{s,2}(z-z_{s})^{2}+\cdots, \\
\eta&=&\eta_{s}+\eta_{s,1}(z-z_{s})+\eta_{s,2}(z-z_{s})^{2}+\cdots,
\end{eqnarray}
around the sonic point.
From equations~(\ref{eq:M'})-(\ref{eq:const}), we find
\begin{eqnarray}
M_{s,1}&=&\frac{k}{1+k}\frac{y_{s}}{1-y_{s}}M_{s}, \\
S_{s,1}&=&-\frac{1-y_{s}}{1+k}S_{s}, \\
\eta_{s,1}&=&-\frac{(3k+4y_{s}-5)\mp\sqrt{D}}{2(1-k)}\eta_{s},
\label{eq:etas1}
\end{eqnarray}
if $D$ is not negative.
It is found that the upper and lower signs correspond to the
$+$ive and $-$ive directions, respectively.
In other words, there are two analytic solutions at a sonic point,
one of which crosses the sonic point 
along the $+$ive direction and the other along 
the $-$ive direction.
Since all coefficients of Taylor series are determined
only by $M_{s}$, $S_{s}$ and $\eta_{s}$, the analytic solution
with each direction is unique for given $y_{s}$.
Hence, for a saddle, two solutions pass it
and both of them are analytic.
On the other hand, for a nondegenerate node, one analytic solution and 
one-parameter family of nonanalytic solutions 
pass it along the primary direction, while
one solution passes it
along the secondary direction and this solution is analytic.
In particular, all nonanalytic solutions must cross nodes
and go along the primary direction.

\section{Kink Instability}
\subsection{Equations for kink mode}
We consider perturbations which satisfy
the following conditions in the background of a self-similar solution:
(1) The initial perturbations vanish inside the sonic point
for $t<0$.
(For $t>0$, the initial perturbations vanish 
outside the sonic point.)
(2) $M$, $S$ and $\eta$ are continuous everywhere, 
in particular at the sonic point.
(3) $\eta^{\prime}$ is discontinuous at the sonic point, 
although it has definite one-sided values as
$z\to z_{s}-0$ and $z\to z_{s}+0$.

We denote the full order perturbations as
\begin{eqnarray}
\delta S(\tau,z)&=&S(\tau,z)-S_{b}(z), \\
\delta M(\tau,z)&=&M(\tau,z)-M_{b}(z), \\
\delta \eta(\tau,z)&=&\eta(\tau,z)-\eta_{b}(z),
\end{eqnarray}
where $M_{b}$, $S_{b}$ and $\eta_{b}$ denote the background
self-similar solution.
Hereafter we often omit the subscript `$b$' unless 
the omission may cause confusion.
By conditions (2) and (3), and 
from equations~(\ref{eq:tauz1}) and (\ref{eq:tauz2}),
it is found that $\delta M'$ and $\delta S'$ are also continuous.
Hence, the perturbations satisfy 
$
\delta M=0,
\delta S=0,
\delta \eta =0,
\delta M'=0,
\delta S'=0,
\delta \eta'\neq 0
$
at the sonic point
at initial moment $\tau=\tau_{0}$,
where the prime denotes the derivative
with respect to $z$ on the perturbed side.
The evolution of the initially unperturbed region
is completely described by the background self-similar solution
because no information from the perturbed side can penetrate
the unperturbed side by condition (1).
Then, by conditions (2) and (3), we find
$
\delta M= 0, 
\delta S= 0, 
\delta \eta= 0, 
\delta \dot{M}=0, 
\delta \dot{S}=0, 
\delta\dot{\eta}=0, 
\delta \dot{M}'=0, 
\delta \dot{S}'=0, 
\delta \dot{\eta}'\neq 0,
$
at the sonic point for $\tau\ge \tau_{0}$ for the case of $t<0$
(for $\tau\le \tau_{0}$ for the case of $t>0$).
This mode has a physical meaning that it injects
discontinuity into the density gradient.
In fact, the density gradient with respect to the physical 
length 
$dl\equiv e^{\omega/2}dr$ on the $t=\mbox{const}$ spacelike
hypersurface is directly related to $\eta'$ as
\begin{equation}
\left.\frac{\partial \rho}{\partial l}\right|_{t}
=\left.e^{-\omega/2}\frac{\partial}{\partial r}\right|_{t}
\frac{\eta}{8\pi r^{2}}=\frac{1}{8 \pi r^{3}}e^{-\omega/2}
(\eta'-2\eta).
\end{equation}

Differentiating equations~(\ref{eq:tauz1}) and (\ref{eq:tauz2})
with respect to $z$ and estimating both sides at the point
of discontinuity, 
we obtain the following equations for the full
order perturbations:
\begin{eqnarray}
\frac{\delta M''}{M}&=&\frac{k}{1+k}\frac{1}{y}\frac{\delta \eta'}{\eta},
\label{eq:dmpp} \\
\frac{\delta S''}{S}&=&-\frac{y}{1+k}\frac{\delta\eta'}{\eta}. 
\label{eq:dspp}
\end{eqnarray}
We also differentiate equation~(\ref{eq:tauz4}) 
with respect to $z$. The result is
\begin{eqnarray}
& &\frac{\ddot{S}'}{S}+2\frac{\dot{S}''}{S}+\frac{S'''}{S}
+\frac{k}{1+k}\left(\frac{\dot{\eta}'}{\eta}-
\frac{\dot{\eta}}{\eta}\frac{\eta'}{\eta}
+\frac{\eta''}{\eta}-\frac{\eta'^{2}}{\eta^{2}}\right)\left(
\frac{\dot{S}}{S}+\frac{S'}{S}\right) \nonumber \\
& &+\left[\frac{k}{1+k}\left(\frac{\dot{\eta}}{\eta}+\frac{\eta'}{\eta}-2
\right)+1\right]\left(\frac{\dot{S}'}{S}+\frac{S''}{S}\right) \nonumber \\
&=&-\frac{k}{1+k}e^{\sigma-\omega-2z}
\left[\left[-\frac{2k}{1+k}\left(\frac{\eta'}{\eta}-2\right)
+\frac{2}{1+k}\frac{\eta'}{\eta}+4\frac{S'}{S}-2\right]
\left(\frac{\eta'}{\eta}-2\right)\left(1+\frac{S'}{S}\right)
\right.\nonumber \\
& &\left.+\left(\frac{\eta''}{\eta}-\frac{\eta'^{2}}{\eta^{2}}\right)
\left(1+\frac{S'}{S}\right)
+\left(\frac{\eta'}{\eta}-2\right)\left(\frac{S'}{S}+\frac{S''}{S}\right)
\right] \nonumber \\
& &-\frac{1}{2}e^{\sigma-2z}\left[\left[-\frac{2k}{1+k}\left(
\frac{\eta'}{\eta}-2\right)-2\right]\left(\frac{M}{S^{3}}+k\eta\right)
\right.\nonumber \\
& &\left.+\left(\frac{M'}{S^{3}}+k\eta'+3k\eta\frac{S'}{S}\right)
-2\frac{S'}{S}\left(\frac{M}{S^{3}}+k\eta\right)\right].
\label{eq:eulerp}
\end{eqnarray}

We find from equations~(\ref{eq:tauz4}) and (\ref{eq:eulerp})
for the full order perturbations at the point of discontinuity
\begin{eqnarray}
& &\frac{\delta S''}{S}+\frac{k}{1+k}\frac{S'}{S}\frac{\delta\eta'}{\eta}=
-\frac{k V^{-2}}{1+k}\left(1+\frac{S'}{S}\right)\frac{\delta\eta'}{\eta}, 
\label{eq:eulerdelta}\\
& &2\frac{\delta \dot{S}''}{S}+\frac{\delta S'''}{S} 
+\frac{k}{1+k}\left(\frac{\delta\dot{\eta}'}{\eta}
+\frac{\delta\eta''}{\eta}-\frac{\delta\eta'^{2}}{\eta^{2}}
-2\frac{\eta'}{\eta}\frac{\delta\eta'}{\eta}
\right)\frac{S'}{S} \nonumber \\
& &+\left[\frac{k}{1+k}\left(\frac{\eta'}{\eta}-2\right)
+1\right]\frac{\delta S''}{S}
+\frac{k}{1+k}\frac{\delta\eta'}{\eta}\frac{S''}{S}+
\frac{k}{1+k}\frac{\delta\eta'}{\eta}
\frac{\delta S''}{S} \nonumber \\
&=&-\frac{k V^{-2}}{1+k}\left[2\frac{1-k}{1+k}\frac{\delta\eta'}{\eta}
\left(\frac{\eta'}{\eta}-2\right)\left(1+\frac{S'}{S}\right)
+2\frac{1-k}{1+k}\left(1+\frac{S'}{S}\right)\frac{\delta\eta'^{2}}
{\eta^{2}}
\right.\nonumber \\
& &+\left[-\frac{2k}{1+k}\left(\frac{\eta'}{\eta}-2\right)
+\frac{2}{1+k}\frac{\eta'}{\eta}
+4\frac{S'}{S}-2\right]\frac{\delta\eta'}{\eta}\left(1+\frac{S'}{S}\right)
\nonumber \\
& &+\left(\frac{\delta\eta''}{\eta}
-2\frac{\eta'}{\eta}\frac{\delta\eta'}{\eta}
-\frac{\delta\eta'^{2}}{\eta^{2}}\right)\left(1+\frac{S'}{S}\right) 
\nonumber \\
& &\left.+\frac{\delta\eta'}{\eta}\left(\frac{S'}{S}+\frac{S''}{S}\right)
+\left(\frac{\eta'}{\eta}-2\right)\frac{\delta S''}{S}
+\frac{\delta\eta'}{\eta}\frac{\delta S''}{S}\right] 
\nonumber \\
& &-\frac{1}{2} e^{\omega}\eta 
V^{-2}\left[-\frac{2k}{1+k}\frac{\delta\eta'}{\eta}
(y+k)+k\frac{\delta\eta'}{\eta}\right].
\label{eq:euler'delta}
\end{eqnarray}
At the sonic point,
equation~(\ref{eq:eulerdelta}) is reduced to
\begin{equation}
\frac{\delta S''}{S}+\frac{y}{1+k}\frac{\delta \eta'}{\eta}=0,
\end{equation}
which is equivalent to equation~(\ref{eq:dspp}).
Therefore, 
it is found that the injection of this type of discontinuity into
the density gradient field is possible 
only at the sonic point.
Next, we evaluate both sides of equation~(\ref{eq:euler'delta})
at the sonic point using 
equations~(\ref{eq:sonic}) and (\ref{eq:regular}).
One can observe that there appear $\delta S'''$ and $\delta\eta''$ 
in this equation.
It is easily found that such terms appear in the equation 
only through the following combination:
\begin{equation}
\frac{\delta S'''}{S}
+\frac{y}{1+k}\frac{\delta \eta''}{\eta}.
\end{equation}
In order to write these terms using the lower order derivatives,
we eliminate $M'$ from equations~(\ref{eq:tauz1}) and 
(\ref{eq:tauz2})
as
\begin{equation}
M-\dot{M}=\eta S^{3}+(1+k)\eta S^{2}S'+k\eta S^{2}\dot{S}.
\end{equation}
Differentiating the above equation twice with respect to $z$, we have
\begin{eqnarray}
& & M''-\dot{M}'' \nonumber \\
&=&\eta''S^{3}+6\eta'S^{2}S'+6\eta SS'^{2}+3\eta S^{2}S''
\nonumber \\
& &+(1+k)(\eta''S^{2}S'+4\eta'SS'^{2}+2\eta'S^{2}S''+2\eta S'^{3}+6\eta SS'S''
+\eta S^{2}S''') \nonumber \\
& &+k(\eta'' S^{2}\dot{S}+4\eta' SS'\dot{S}+2\eta'S^{2}\dot{S}'
+2\eta S'^{2}\dot{S}+2\eta SS''\dot{S}+4\eta SS'\dot{S}'
+\eta S^{2}\dot{S}'').
\end{eqnarray}
Estimating the full order perturbations of 
both sides at the sonic point, we find
\begin{eqnarray}
& &\frac{\delta S'''}{S}+\frac{y}{1+k}\frac{\delta \eta''}{\eta} 
\nonumber \\
&=&\frac{y}{1+k}\left[\frac{\delta M''}{M}-\frac{\delta \dot{M}''}{M}\right]
-\frac{k}{1+k}\frac{\delta \dot{S}''}{S}
-2\left[\frac{3}{1+k}\frac{S'}{S}+2\left(\frac{S'}{S}\right)^{2}
+\frac{S''}{S}\right]\frac{\delta \eta'}{\eta}\nonumber \\
& &-\left[\frac{3}{1+k}+2\left(\frac{\eta'}{\eta}+3\frac{S'}{S}\right)
\right]\frac{\delta S''}{S} 
-2\frac{\delta \eta'}{\eta}\frac{\delta S''}{S}.
\end{eqnarray}
Using the above relation, we evaluate both sides of 
equation~(\ref{eq:euler'delta}) at the sonic point
using $\delta M''$, $\delta S''$ and $\delta \eta'$.
After a rather lengthy calculation, 
we obtain the following equation:
\begin{equation}
\frac{\delta\dot{\eta}'}{\eta}
+\left(-2\frac{1-k}{1+k}\frac{\eta'}{\eta}+\frac{5-3k-4y}{1+k}\right)
\frac{\delta\eta'}{\eta}
-\frac{1-k}{1+k}\left(\frac{\delta\eta'}{\eta}\right)^{2}=0.
\end{equation}
It is noted that the PDE's become an ODE
for the discontinuity in the
density gradient at the sonic point.
This is due to the nature of the sonic point.
It is more convenient to rewrite this in terms of
$V'$ and $\delta V'$ using the relation (\ref{eq:v'eta'}).
The discontinuity in $V'$ is directly related to the discontinuity 
in $\delta\eta'$ through
\begin{equation}
\frac{\delta V'}{V}=-\frac{1-k}{1+k}\frac{\delta \eta'}{\eta}.
\label{eq:deltav'deltaeta'}
\end{equation}
Finally we obtain the following equation for the full order
perturbation as
\begin{equation}
\frac{\delta\dot{V}'}{V}
-\left[1-2\left(\frac{V'}{V}\right)_{b}\right]
\frac{\delta V'}{V}
+\left(\frac{\delta V'}{V}\right)^{2}=0.
\label{eq:master}
\end{equation}
It is noted that the above equation has complete correspondence to
its Newtonian counterpart,
equation~(15) of~\cite{op1988}, when the latter is rewritten 
in terms of the relative velocity of the constant self-similar coordinate 
curve to the fluid element.

\subsection{Linear order analysis}
By linearizing equation~(\ref{eq:master}), we can obtain
\begin{equation}
\frac{\delta\dot{V}'}{V}=
\left[1-2\left(\frac{V'}{V}\right)_{b}\right]\frac{\delta V'}{V}.
\end{equation}
This equation is integrated as
\begin{equation}
\frac{\delta V'}{V}=\mbox{const}\cdot e^{\alpha\tau},
\end{equation}
where
\begin{equation}
\alpha\equiv 1-2\left(\frac{V'}{V}\right)_{b}.
\end{equation}
Therefore, it is found that the discontinuity 
in $V'$ grows for $\alpha>0$, decays
for $\alpha<0$ and 
the mode is neutral for $\alpha=0$ as $\tau$ increases.
On the other hand, it is found that the discontinuity 
in $V'$ grows for $\alpha<0$, decays
for $\alpha>0$ and
the mode is neutral for $\alpha=0$ as $\tau$ decreases.
The above discussion will suggest some kind of instability.
Nevertheless, the linear order analysis will not be sufficient
in this case from the following reason. The discontinuity in
the density gradient which is injected as a perturbation
may not be included in the background solution.
Then, there is no natural measure
how much the perturbation grows compared with the background
solution.
In the linear order analysis, we can show at most a power-law
growth of the discontinuity
for sufficiently small perturbation. In the next subsection,
we will show a divergence of full order perturbation
at some finite moment, even if 
initial perturbation is small.
This does indicate instability.

\subsection{Full order analysis}
In fact, equation~(\ref{eq:master}) is easily integrated. Putting
\begin{eqnarray}
X&\equiv &\frac{\delta V'}{V},
\end{eqnarray}
we rewrite equation~(\ref{eq:master}) as
\begin{equation}
\dot{X}-\alpha X+X^{2}=0.
\end{equation}
There are two stationary solutions
$X=0$ and $X=\alpha$.
General solutions are the following:
\begin{eqnarray}
X&=&\frac{\alpha}
{1-\exp(-\alpha\tau+\mbox{const})} , \quad\mbox{for}\quad \alpha\ne 0, \\
X&=&\frac{1}{\tau+\mbox{const}}, \quad\mbox{for}\quad \alpha=0.
\end{eqnarray} 
For an initial value
$X=X_{0}\ne 0$ at $\tau=\tau_{0}$, the solution is written as
\begin{eqnarray}
\frac{1}{X}&=&\frac{1}{\alpha}-e^{-\alpha(\tau-\tau_{0})}
\left(\frac{1}{\alpha}-\frac{1}{X_{0}}\right),
\quad \mbox{for}\quad 
\alpha\ne 0, 
\label{eq:InvAne0}\\
\frac{1}{X}&=&\frac{1}{X_{0}}+(\tau-\tau_{0}), \quad \mbox{for}
\quad \alpha=0.
\label{eq:InvA0}
\end{eqnarray}

\subsubsection{$D>0$ case}
First, we consider $\alpha>0$.
From equation~(\ref{eq:InvAne0}), it is seen 
that, as $\tau$ increases, $X$ blows up to $-\infty$
at some finite moment $\tau\to \tau_{d}-0>\tau_{0}$ for $X_{0}<0$, where
\begin{equation}
\tau_{d}\equiv\tau_{0}+\frac{1}{\alpha}
\ln\left(1-\frac{\alpha}{X_{0}}\right),
\end{equation}
while $X$ monotonically 
approaches $\alpha $ for $X_{0}>0$. 
Next, we consider $\alpha<0$.
From equation~(\ref{eq:InvAne0}), it is seen that,
as $\tau$ increases, 
$X$ monotonically 
approaches $0$ for $X_{0}>\alpha$, while $X$ blows up to $-\infty$
at some finite moment $\tau=\tau_{d}-0>\tau_{0}$
for $X_{0}<\alpha$.

Actually these two analyses under two different backgrounds are
only apparently different local pictures of
nonlinear dynamics.
If we consider the background solution which
crosses the sonic point along the $+$ive direction,
$\alpha$ is positive and the stationary solution 
$X=\alpha$ corresponds to 
another possible direction, since
\begin{equation}
\left(\frac{V'}{V}\right)_{-}=
\left(\frac{V'}{V}\right)_{+}
+\left[1-2\left(\frac{V'}{V}\right)_{+}\right].
\end{equation}
The situation is reversed for the $-$ive direction solution, since
\begin{equation}
\left(\frac{V'}{V}\right)_{+}=
\left(\frac{V'}{V}\right)_{-}+\left[1-2
\left(\frac{V'}{V}\right)_{-}\right].
\end{equation}

\begin{figure}[htbp]
\centerline{
\epsfxsize 15cm \epsfysize 15cm
\epsfbox{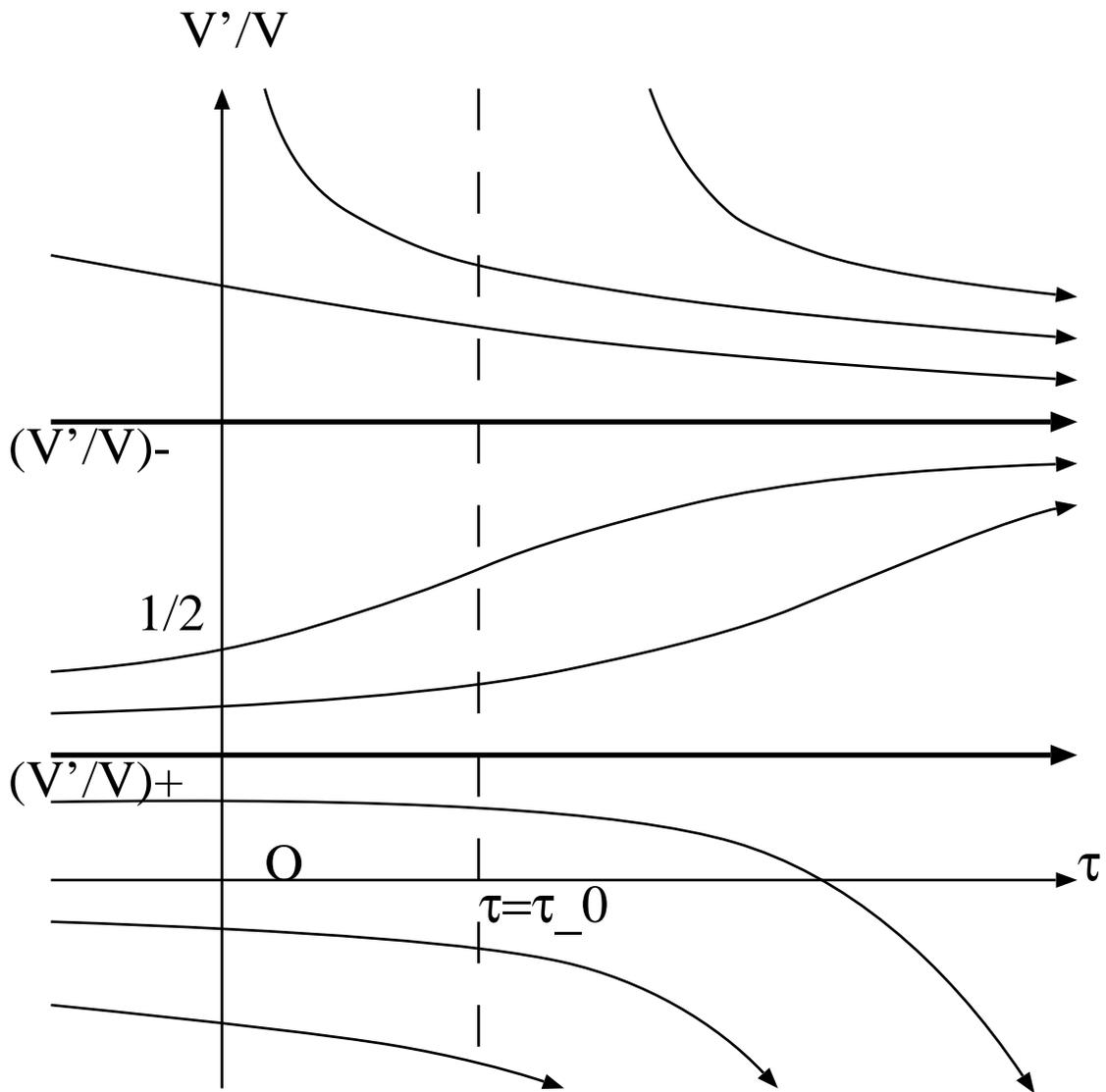}}
\caption{Dynamical behavior of the kink mode for a 
nondegenerate-nodal sonic point.
The figure is the same except for $(V'/V)_{+}< 0$ for a saddle
sonic point.}
\label{fg:Dpositive}
\end{figure}

The global picture is the following.
With the initial value $V'/V=(V'/V)_{0}$ at $\tau=\tau_{0}$,
we obtain the following full order behavior of the perturbation:
\begin{eqnarray}
\frac{V'}{V}&\to& -\infty \quad \mbox{as}\quad \tau\to \tau_{d}-0
\quad \mbox{for}\quad \left(\frac{V'}{V}\right)_{0}<
\left(\frac{V'}{V}\right)_{+}, \\
\frac{V'}{V}&=& \left(\frac{V'}{V}\right)_{+}
=\mbox{const}\quad \mbox{for}\quad 
\left(\frac{V'}{V}\right)_{0}=
\left(\frac{V'}{V}\right)_{+}, \\
\frac{V'}{V}&\to& \left(\frac{V'}{V}\right)_{-}\quad 
\mbox{as}\quad 
\tau\to \infty \quad \mbox{for}\quad \left(\frac{V'}{V}\right)_{+}<
\left(\frac{V'}{V}\right)_{0},
\end{eqnarray}
as $\tau$ increases, while
\begin{eqnarray}
\frac{V'}{V}&\to& \left(\frac{V'}{V}\right)_{+} 
\quad \mbox{as}\quad \tau\to -\infty
\quad \mbox{for}\quad \left(\frac{V'}{V}\right)_{0}<
\left(\frac{V'}{V}\right)_{-}, \\
\frac{V'}{V}&=& \left(\frac{V'}{V}\right)_{-}
=\mbox{const}\quad \mbox{for}\quad 
\left(\frac{V'}{V}\right)_{0}=
\left(\frac{V'}{V}\right)_{-}, \\
\frac{V'}{V}&\to& \infty\quad 
\mbox{as}\quad \tau\to \tau_{d}+0 \quad \mbox{for}\quad 
\left(\frac{V'}{V}\right)_{-}<
\left(\frac{V'}{V}\right)_{0},
\end{eqnarray}
as $\tau$ decreases.
The dynamical behavior of the perturbation is schematically
depicted in figure~\ref{fg:Dpositive}.

\subsubsection{$D=0$ case}
We consider $\alpha=0$.
From equation~(\ref{eq:InvA0}), it is seen that, as $\tau$ increases, 
$X$ diverges to $-\infty$ 
at some finite moment $\tau\to\tau_{d}$, where 
\begin{equation}
\tau_{d}\equiv \tau_{0}-\frac{1}{X_{0}},
\end{equation}
for $X_{0}<0$, while 
$X$ approaches $0$ for $X_{0}>0$.

\begin{figure}[htbp]
\centerline{
\epsfxsize 15cm \epsfysize 15cm
\epsfbox{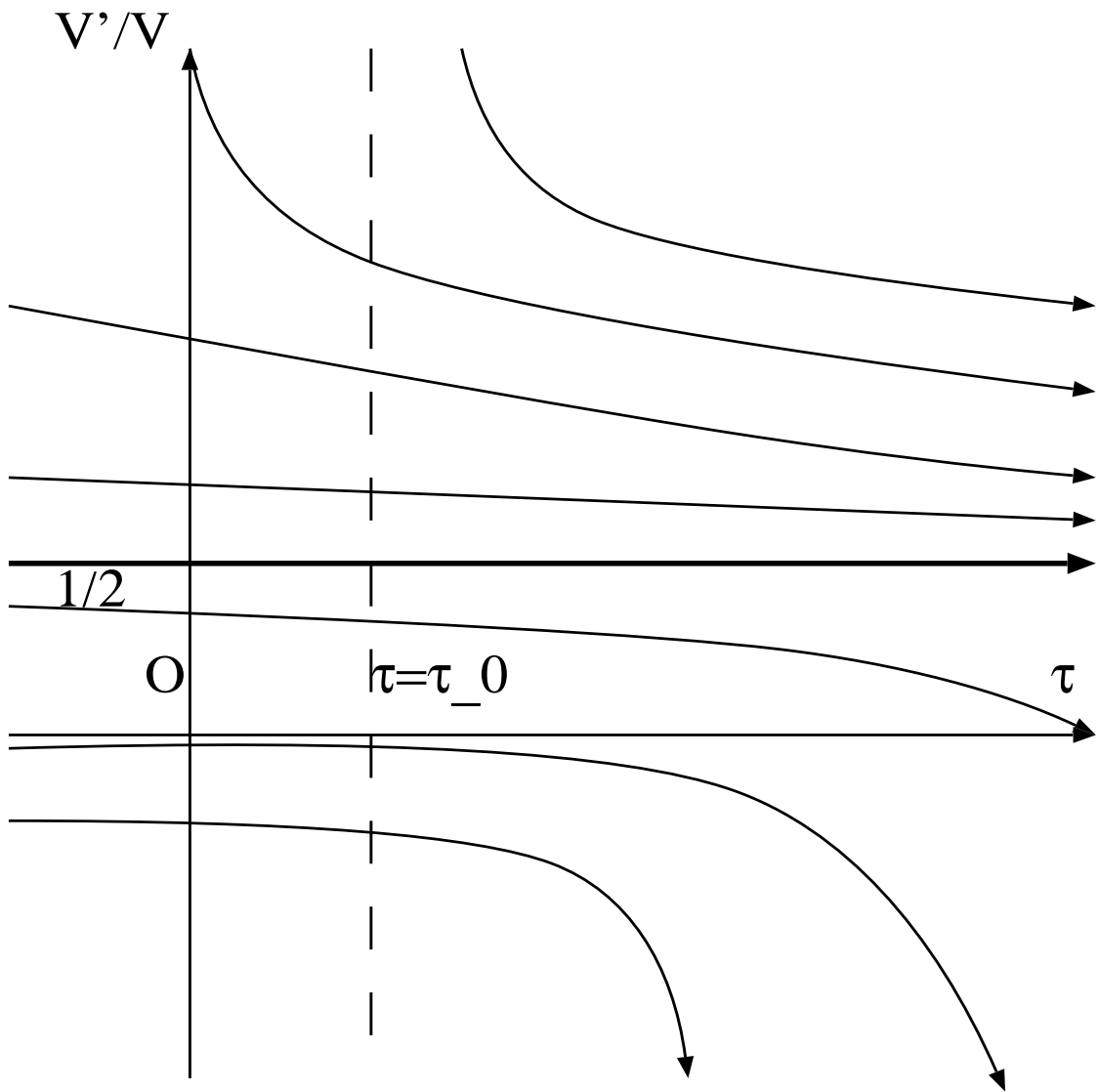}}
\caption{Dynamical behavior of the kink mode for a degenerate-nodal
sonic point.}
\label{fg:Dzero}
\end{figure}

This case corresponds to a degenerate node for which 
$D=0$ and $(V'/V)_{b}=1/2$.
Then we find
\begin{eqnarray} 
\frac{V'}{V}&\to& -\infty \quad \mbox{as}\quad \tau\to \tau_{d}-0
\quad \mbox{for}\quad 
\left(\frac{V'}{V}\right)_{0}<\frac{1}{2}, \\
\frac{V'}{V}&\to& \frac{1}{2}
\quad \mbox{as}\quad 
\tau\to \infty \quad \mbox{for}\quad \frac{1}{2}\le 
\left(\frac{V'}{V}\right)_{0},
\end{eqnarray}
as $\tau$ increases, while
\begin{eqnarray} 
\frac{V'}{V}&\to& \frac{1}{2} \quad \mbox{as}\quad \tau\to -\infty
\quad \mbox{for}\quad 
\left(\frac{V'}{V}\right)_{0}\le \frac{1}{2}, \\
\frac{V'}{V}&\to& \infty
\quad \mbox{as}\quad 
\tau\to \tau_{d}+0 \quad \mbox{for}\quad \frac{1}{2} <
\left(\frac{V'}{V}\right)_{0},
\end{eqnarray}
as $\tau$ decreases.
The behavior is depicted in figure~\ref{fg:Dzero}.

\subsection{Stability criterion}
First, we consider $t<0$. The time evolution is regarded
as the increase of $\tau$ from $\tau=\tau_{0}$.
Then, we obtain the following criterion:
all $+$ive direction solutions and degenerate-nodal-point ones
are unstable, while all $-$-direction solutions
are stable for this mode.
Next, we consider the case $t>0$. In that case, the time evolution
is regarded as the decrease of $\tau$
from $\tau=\tau_{0}$. 
Then, we find the following criterion:
all $-$-direction solutions and degenerate-nodal-point ones are
unstable, while all $+$-direction solutions
are stable for this mode.
These are summarized in table~\ref{tb:class}.

The stability criterion which applies to all possible situations
is the following:
if $V'\ge V/2$ at the sonic point, then the solution is unstable,
while, if not, then the solution is stable
for the kink mode.    
This criterion can be rewritten in terms of the partial derivative 
with respect to $t$ or $r$ as follows:
in terms of the derivative with respect to $t$,
if $\partial V^{2}/\partial t + k/t\ge 0$, then the solution is unstable,
while, if not, the solution is stable for the kink mode;
in terms of the partial derivative with respect to $r$, 
if $\partial V/\partial r\ge V/(2r)$ at the sonic point,
the solution is unstable, while, if not,
the solution is stable for this mode. 

However, it should be noted that 
we cannot say that a solution is stable because our
analysis is specified on the kink mode. 
The present criterion for stability should be considered
as a necessary condition for stability, while 
the criterion for instability should be considered as
a sufficient condition for instability.

\section{Applications}
\subsection{Flat Friedmann solution}
The flat Friedmann solution is a member of self-similar solutions.
This solution is characterized by $y=1/3$ and $\eta'/\eta=2$.
This solution crosses a transonic point.
For $y_{s}=1/3$, we find two nonzero eigenvalues
from equation~(\ref{eq:tildelambda}) as 
\begin{equation}
\tilde{\lambda}=-\frac{2}{3}, -k-\frac{1}{3}.
\end{equation}
Therefore, the sonic point is a nondegenerate node
for $k\ne 1/3$ and a degenerate node for $k=1/3$. 
Since we find from equation~(\ref{eq:eta's})
\begin{equation}
\tilde{\lambda}=-k-\frac{1}{3},
\end{equation}
the flat Friedmann solution
crosses a nondegenerate
node along the primary direction for $0<k<1/3$,
a degenerate node for $k=1/3$, 
and a nondegenerate node along the secondary direction for $1/3<k<1$. 
The solution is expanding for $t>0$, while
the solution is collapsing for $t<0$.
From the obtained stability criterion,
we find that the expanding flat Friedmann solution
is unstable against the kink mode for $1/3\le k<1$,
which implies that it may not be a good cosmological model
for $1/3\le k<1$.
We find that the collapsing flat Friedmann solution is also unstable for 
$0<k\le 1/3$ against the kink mode,
which implies that generic gravitational collapse
does not proceed homogeneously for $0<k\le 1/3$.
We also find that the expanding flat Friedmann
solution for $0<k<1/3$ and the collapsing
flat Friedmann solution for $1/3<k<1$ 
do not suffer kink instability.

\subsection{Static self-similar solution}
The static self-similar solution is 
characterized by $y=1$ and $\eta'/\eta=0$.
This solution crosses a transonic point.
For $y_{s}=1$, we find eigenvalues from equation~(\ref{eq:tildelambda}):
\begin{equation}
\tilde{\lambda}=-2k, -1+k.
\end{equation}
Since we find from equation~(\ref{eq:eta's})
\begin{equation}
\tilde{\lambda}=-1+k,
\end{equation}
the static self-similar solution
crosses a nondegenerate node along the secondary direction for $0<k<1/3$,
a degenerate node for $k=1/3$, 
and a nondegenerate node along the primary direction for $1/3<k<1$. 
From the stability criterion for the kink mode,
we find that the solution
is unstable for $t>0$ for $0<k<1/3$.
We also find that the solution
is unstable for $t<0$ for $1/3<k<1$.
For $k=1/3$, the solution is unstable, irrespective of the sign of $t$.
However, it should be noted that the moment $t=0$ has 
no physical meaning in the static solution
because the system is invariant under time translation.
Therefore, we conclude that the static self-similar solution
is unstable for $0<k<1$.

\subsection{Nonanalytic self-similar solutions}
As we have seen, there are one-parameter family of nonanalytic 
self-similar solutions which cross a sonic point.
All of them cross nodes along the primary direction.
When the stability criterion is applied,
these nonanalytic solutions are all unstable for $t<0$.
Thereby, stable self-similar collapse solutions with sonic points
must be analytic.
If we consider self-similar solutions with regular 
center and sonic points,
a set of all stable collapse solutions
becomes discrete,
while a set of all collapse solutions
is dense.
In other words, the stability requirement for the kink mode 
considerably reduces the number of self-similar collapse solutions.
For $t>0$, nonanalytic self-similar solutions do not suffer
kink instability.

\subsection{Anti-transonic point}
We have seen that an anti-transonic point must be a saddle
and that the solution crosses it along the $+$ive direction.
From the stability criterion,
the solution is unstable for $t<0$.
Therefore, in collapse, any stable self-similar solution  
cannot pass an anti-transonic point.

\subsection{Larson-Penston (attractor) solution}
The Larson-Penston solution is a solution which is 
analytic both at the center and at a transonic point
and it has no zero
in the velocity field $V_{R}$. 
The Larson-Penston solution has no
analytic unstable mode in Newtonian 
gravity~\cite{hn1997hm2000,mh2001}
and in general relativity 
at least for $0<k\le 0.03$~\cite{hm2001}.
The nature of the sonic point was studied 
by~\cite{op1990,ccgnu2000}.

This solution crosses the sonic point
along the secondary direction for $0<k\alt 0.036$.
It has been found that the sequence of these analytic solutions
changes from the secondary-direction nodal-point solution
to the degenerate-nodal-point one at $k\simeq 0.036$.
For a larger value of $k$, the sequence 
changes from the degenerate-nodal-point solution
to the primary-direction nodal-point one~\footnote{
Although we could not find that in literature, 
we have confirmed it numerically.}.
By the present stability criterion for $t<0$, we conclude that 
the Larson-Penston solution is stable for the kink mode
for $0<k\alt 0.036$, while
it is unstable for $0.036\alt k$.
It is important that the kink mode does not affect
the nature of the Larson-Penston solution as an attractor 
for $0<k\alt 0.036$, while it does for $0.036\alt k$.

For $t>0$, the time-reversed Larson-Penston
solution is unstable against the kink mode
for $0<k\alt 0.036$, while it is stable for $0.036\alt k$
for this mode.
This suggests that the solution cannot describe
a realistic expanding inhomogeneous universe for $0<k\alt 0.036$.
From the Larson-Penston solution for $t<0$, there is another
method to construct the solution for $t>0$.
That is the analytic continuation of the solution
for $t<0$ to $t>0$.
Since the Larson-Penston solution is `quasi-static',
the analytic continuation beyond $\xi=0$ is 
possible~\cite{cc2000,op1987,op1990,fh1993}.
This continuated solution for $t>0$ does not cross
a sonic point~\cite{op1987,op1990,fh1993}.
Therefore, the solution has no kink mode for $t>0$.

\subsection{Evans-Coleman (critical) solution}
Evans-Coleman solution is a solution which is analytic
both at the center and a transonic point and it has
one zero in the velocity field $V_{R}$.
The Evans-Coleman solution has a single analytic unstable mode,
which was shown for 
$0<k\le 0.889$ by normal mode analyses~\cite{kha1995,maison1996kha1999}
and for $0<k\le 1$ by numerical simulations~\cite{nc2000}.
The nature of a sonic point and the asymptotic form
of this solution were studied by~\cite{nc2000,ccgnu2000}.

The sequence of these analytic solutions
changes its character for $k\simeq 0.41$
and $k\simeq 0.89$ in terms of the sonic point.
The solution crosses a saddle sonic point
for $0<k\alt 0.41$.
For $0.41\alt k \alt 0.89$, the solution crosses a nodal sonic point
in the secondary direction, a degenerate-nodal sonic point 
for $k\simeq 0.89$,
and a nodal sonic point in the primary direction for 
$0.89\alt k$.
From the stability criterion,
it is found that the kink mode does not affect the critical
nature of the Evans-Coleman solution for $0<k\alt 0.89$,
while it is also found that the Evans-Coleman 
solution suffers the kink instability for $0.89\alt k$.
Since the critical solution is assumed to have a single unstable 
mode, it indicates that the Evans-Coleman solution
for $0.89\alt k$ cannot be a critical solution
because the solution has one analytic unstable mode
and one non-analytic unstable mode, i.e., the kink mode.

For $t>0$, 
the time-reversed Evans-Coleman solution is unstable for $0<k\alt 0.89$,
while it is stable for $0.89\alt k$ for the kink mode. 
This suggests that the solution would not be a good model of 
the expanding inhomogeneous universe for $0<k\alt 0.89$.
The behavior of the solution as $\xi$ increases from $-\infty$
changes at $k\simeq 0.28$ from the quasi-static solution
to the `asymptotically Minkowski' solution as $k$ increases~\cite{ccgnu2000}.
Only for the quasi-static solution, 
the analytic continuation
beyond $\xi=0$ is possible.
However, it was shown that the analytically continuated 
Evans-Coleman solution
beyond $\xi=0$ encounters
another sonic point and will not be able to cross it
regularly~\cite{op1990,ccgnu2000,fh1993}.

\section{Discussions}
\subsection{Comment on Neilsen and Choptuik~\protect\cite{nc2000}}
Here we should comment on the recent work by Neilsen and
Choptuik~\cite{nc2000}.
They numerically simulated the PDE's (Einstein's equations 
and equations of motion for a perfect fluid)
and found critical behavior
even for $0.89\alt k$.
They found that the critical solution obtained by solving 
the PDE's well agreed with the Evans-Coleman self-similar solution
which is obtained by solving the ODE's with the requirement of
analyticity at the sonic point.
By numerical simulations of the PDE's,
they measured the critical exponent of the power law
which the formed black hole mass obeys.
The obtained value of critical exponent is 
continuous with respect to the change of $k$.
It implies the existence of one analytic unstable mode. 
In contrast, we have shown that the Evans-Coleman
solution suffers kink instability for $0.89\alt k$.
We would like to point out the following three possibilities.
The first is that perturbations which belonged to
the kink mode in their numerical simulations 
might have been so small that there 
would have been little time 
for the kink mode to grow up sufficiently until the 
numerical simulations were stopped for other reasons.
The second is that some kind of numerical viscosity,
which is usually included in numerical codes of fluid dynamics
deliberately or not deliberately,
might have killed the kink instability.
The third possibility is that the nonlinear coupling of the 
kink mode with other analytic modes might have weakened 
the growth of instability.
In spite of such possibilities, 
for $0.89\alt k$, the Evans-Coleman has
two unstable modes, one is analytic and the other is not analytic.

\subsection{Two Limiting cases}
It is remarkable that
the stability analysis for the kink mode of self-similar solutions
in general relativity is completely
parallel to that in Newtonian gravity.
The stability for the kink mode is determined by the
class of the pertinent sonic point.
Therefore, it is clear that the present analysis on the stability
of general relativistic self-similar solutions 
contains the Newtonian analysis for isothermal gas 
as a limit of $k\to 0$.
On the other hand, it should be noted that $k=0$, i.e., a dust fluid, 
is not included in the present analysis since
self-similar solutions with a dust fluid 
cannot be regarded as a continuous limit of $k\to 0$
of those with $P=k\rho$ 
in several respects~\cite{op1990}.

Since we have concentrated on self-similar solutions
with $P=k\rho$ for $0<k<1$, 
we have not dealt with a `stiff matter', i.e., $k=1$. 
This is because, as seen 
in equations~(\ref{eq:zs})-(\ref{eq:etas}), 
(\ref{eq:eta's}), (\ref{eq:v'eta'}), (\ref{eq:etas1}) and 
(\ref{eq:deltav'deltaeta'}),
it is clear that more careful treatment is needed
for that case. The stiff matter is one of the most important 
examples that can be described by the perfect fluid model because
it is equivalent to a massless scalar field under certain
conditions~\cite{madsen1988cg1999}. 
In spite of this equivalence, the critical phenomena
which have been observed in these two systems
look very different~\cite{choptuik1993,nc2000}.
Hence, the analysis for a stiff matter will be 
very interesting.

\section{Summary}
We have studied the stability of self-similar solutions
with perfect fluids in general relativity.
It has been found that a wide class of self-similar solutions
turn out to be unstable against the kink mode.
The development of this instability 
will result in the formation of a shock wave.
Since the present analysis is specified only on
the kink mode, we cannot say about generic 
outcome of the growth of this instability.
However, it is
probable that the nonlinear coupling with analytic modes
may play important roles in the development of 
the instability.
The proposition that the stability of solutions for this mode
is solely governed by the nature of the pertinent sonic point
applies to general relativity as well as to Newtonian gravity.

The obtained criterion has been applied to several known
self-similar solutions in general relativity.
Then, it turns out that the expanding flat Friedmann solution
has a growing kink mode for $1/3\le k<1$.
The existence of the growing kink mode in the flat Friedmann solution
with a radiation fluid might affect the standard cosmological
structure formation scenario.
We have also investigated the stability of self-similar collapse 
solutions which have recently called attention in the 
convergence and critical phenomena
in the gravitational collapse of a 
perfect fluid in general relativity.
The results are that the Larson-Penston (attractor) 
solution loses its attractive
nature for $0.036\alt k$ and that the Evans-Coleman 
(critical) solution loses its critical nature 
for $0.89\alt k$.

\acknowledgments

I am grateful to K~Nakao, T~Koike and H~Maeda for helpful discussions.
I would like to thank K~Maeda for 
continuous encouragement.
This work was supported by the 
Grant-in-Aid for Scientific Research (No. 05540)
from the Japanese Ministry of
Education, Culture, Sports, Science and Technology.

\begin{table}[htbp]
  \caption{Stability for the kink mode and the class of sonic points}
  \label{tb:class}
\begin{center}
    \begin{tabular}{c|ccccc} 
      $t$  & Primary node& Secondary node & Degenerate node 
      & Saddle: $+$ $^{\dagger}$& Saddle: $-$ \\ \hline
      $>0$& Stable & Unstable & Unstable & Stable & Unstable  \\ 
      $<0$& Unstable & Stable & Unstable & Unstable & Stable
    \end{tabular}
  \end{center}
{\footnotesize
$^{\dagger}$ No transonic 
point belongs to this class, 
while all anti-transonic points belong to this class.}
\end{table}

\end{document}